# Adaptive Heart Rate Estimation from Face Videos

Utkarsh SHARMA    Terumi UMEMATSU    Masanori TSUJIKAWA    Yoshifumi ONISHI

Biometrics Research Laboratories, NEC Corporation, Japan

**Abstract—** We propose a novel heart rate (HR) estimation method from facial videos that dynamically adapts the HR pulse extraction algorithm to separately deal with noise from 'rigid' head motion and 'non-rigid' facial expression. We first identify the noise type, based on which, we apply specific noise removal steps. Experiments performed on popular database show that the proposed method reduces HR estimation error by over 32%.

1. Introduction

Non-contact Heart Rate (HR) estimation, predominantly from face videos, is increasingly being seen as a non-intrusive, flexible way of determination of a person's physical and mental health. Several methods have been tried in the past to extract cardiac activity-induced color variations on the face [1]-[3], but they are often vulnerable to various noise. These methods use a common noise-removal method to handle these fundamentally different noises. On the contrary, we separately deal with 'rigid noise' caused by head motions and 'non- rigid noise' caused by facial expressions as major noise types.

We propose a novel HR estimation method that detects the kind of noise present in each frame and adapts the HR pulse extraction method to include specific noise removal steps, in order to remove rigid and non-rigid noise separately and selectively.

2. Proposed Method

In the proposed method, first, the facial feature point on the nose and those around the lips are observed over time to detect head motion and facial expressions, respectively, and one of three labels *'motion'*/*'expression'*/*'still'* is assigned to each frame. Fig. 1 shows the block diagram of proposed method. After frame labeling, a rectangular region of interest (ROI) is selected on the face and divided into $N$ sub-ROIs. For each sub-ROI, $r = 1,..., N$, the observed signal $G_r(t)$, green channel mean over all pixels, is computed.

For frames labeled *'still'*, i.e. noise-less, or *'motion'*, reliable HR components are emphasized using an ROI filter that assigns weights $w_r$, inversely proportional to local variance of $G_r(t)$ in a short-time window. For *'expression'* frames, due to complex movement of facial muscles, there is high amplitude noise which inflates local variance of most sub-ROI signals. So, to suppress high amplitude noise, ROI filter weights $v_r$ are assigned inversely proportional to maximum value of $G_r(t)$ in the short-time window. Besides label-dependent ROI filter, label-dependent noise correction steps are applied.

In (2), head motion signal $h(t)$, which adds noise in observed signal $G_r(t)$, is subtracted from $G_r(t)$. Here, α is computed to minimize $\|G_r(t) - \alpha h(t)\|^2$.

*Still:* $\quad G(t) = \Sigma_r w_r G_r(t)$ (1)

*Motion:* $\quad G(t) = \Sigma_r w_r (G_r(t) - \alpha * h(t))$ (2)

*Expression:* $\quad G(t) = \Sigma_r v_r (G_r(t)/\max_t(x_r(t)))$ (3)

In (3), $G_r(t)$ is further weighed by $x_r(t)$, the movement of facial feature point inside/nearest to sub-ROI r, which serves to suppress sub-ROIs with large facial muscle movement (i.e. noise).

Finally, cleaned signal $G(t)$, containing emphasized HR component, is analyzed in frequency domain to estimate HR, generally the frequency corresponding to the highest FFT peak. However, to remove residual noise peaks in presence of severe rigid/non-rigid motion, a peak tracking approach is taken. FFT peaks are tracked over time and in noisy (*'motion'*/ *'expression'*) periods, instead of simply choosing the highest FFT peak, we choose the FFT peak closest to the HR obtained on neighboring *'still'* frames (more reliable since no noise).

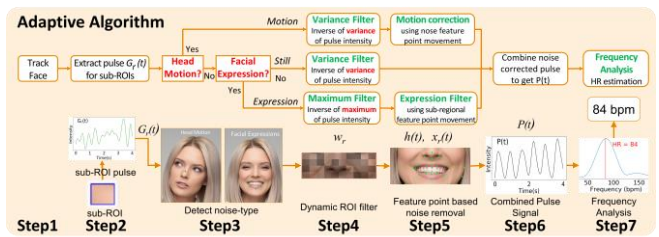

Figure 1: Block diagram and process flow of proposed adaptive heart rate estimation algorithm

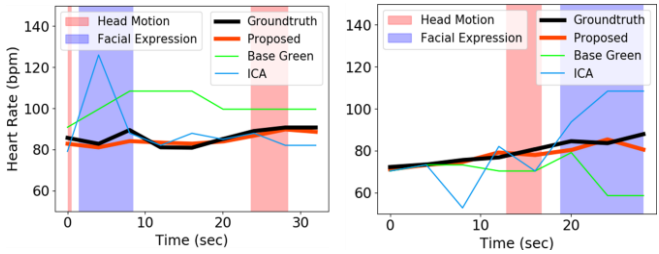

Figure 2: Results show that proposed method (red) is more accurate, especially in the presence of noise

3. Experimental Results

We evaluated the proposed method on a benchmark data set called MMSE-HR [4], a set of 102 videos of 40 subjects (17 male and 23 female). We set the number of sub-ROIs, N = 40 and estimated HR by extracting HR pulse signal (as per section II) over 4 second-long (100 frames) non-overlapping windows. As we can see in Fig. 2, the proposed method (e) clearly outperforms the baseline methods, especially in the presence of 'rigid' and 'non-rigid' noise. As shown in Table I, Root Mean Square Error (RMSE) and Standard Deviation of Error (SDE) with respect to ground truth HR values, were calculated for the proposed method along with previous works described in [1]-[3]. Method (d) is the proposed method without peak tracking, and method (e) is the proposed method including the peak tracking step.

TABLE I. Short-Time HR Estimation Error On Face Videos

| *Methods* | *RMSE** | *SDE** |
|---|---|---|
| (a) Green channel base [1] | 21.68 | 12.62 |
| (b) Independent Component Analysis [2] | 19.92 | 13.39 |
| (c) Self Adaptive Matrix Completion [3]** | 11.66 | 11.51 |
| (d) Proposed: Adaptive HR estimation | 8.95 | 6.10 |
| **(e) Proposed: (d) + peak tracking** | **7.82** | **5.08** |

*RMSE and SDE are in beats-per-minute (bpm)
**These error values are referred from [3]

4. Conclusion

The proposed method (e) achieves the lowest error by adapting the pulse extraction and noise removal steps, based on whether 'rigid' or 'non-rigid' motion is present in the video frames. In the future, we will further analyze cases which involve rigid and non-rigid motion occurring at the same time, although, here, we treated the dominant noise exclusively.